\documentclass[a4paper,11pt]{article}
\pdfoutput=1 

\usepackage{jheppub} 

\usepackage[T1]{fontenc} 
\usepackage{tensor}
\newcommand{\bs}{\begin{split}}
\newcommand{\es}{\end{split}}
\newcommand{\mc}{\mathcal}
\newcommand{\ks}{$Kerr$-$Sen~$}
\newcommand{\eqr}{\eqref}
\newcommand{\A}{\mc{A}}
\newcommand{\Q}{\mc{Q}}
\newcommand{\neks}{$NHNEKS~$}
\title{\boldmath Near-Extremal Black Hole Thermodynamics from $AdS_2/CFT_1$ Correspondence in The Low Energy Limit of 4D Heterotic String Theory}

\author[a]{Bradly K. Button,}
\author[b]{Leo Rodriguez}
\author[a,b]{and Sujeev Wickramasekara}

\affiliation[a]{Department of Physics and Astronomy\\The University of Iowa\\Iowa City, IA 52242}
\affiliation[b]{Department of Physics\\Grinnell College\\Grinnell, IA 50112}

\emailAdd{bradly-button@uiowa.edu}
\emailAdd{rodrigul@grinnell.edu}
\emailAdd{wickrama@grinnell.edu}

\abstract{We compute the full asymptotic symmetry group of the four dimensional near-extremal $Kerr$-$Sen$ black hole within an $AdS_2/CFT_1$ correspondence. We do this by performing a Robinson-Wilczek two dimensional reduction and construct an effective quantum theory of the remaining field content. The resulting energy momentum tensor generates an asymptotic Virasoro algebra, to $s$-wave, with a calculable central extension. This center in conjunction with the proper regularized lowest Virasoro eigen-mode yields the near-extremal $Kerr$-$Sen$ entropy via the statistical Cardy formula. Finally we analyze quantum holomorphic fluxes of the dual $CFT$ giving rise to a finite Hawking temperature weighted by the central charge of the near-extremal $Kerr$-$Sen$ metric.}

\begin{document} 
\maketitle
\flushbottom
\section{Introduction}
\label{sec:intro}
Since the seminal discovery of Brown and Henneaux \cite{brownhenau}, the idea that most black holes are holographically dual to a lower dimensional $CFT$ has become a universal concept \cite{strom2,Carlip:2011ax,carlip,carlip3,carlip2,Park:1999tj,Park:2001zn,kkp,Majhi:2011ws,Majhi:2012tf,Majhi:2012st,Majhi:2012nq}. This concept is most notably exemplified by the $Kerr/CFT$ correspondence \cite{kerrcft} and its extensions \cite{Compere:2012jk,SheikhJabbaria:2011gc,deBoer:2011zt,Yavartanoo2012410,springerlink:10.1140/epjc/s10052-012-1911-7,rasmussen:2010xd,rasmussen:2010sa,Chen:2010yu,Chen:2010ni,Chen:2010xu,Chen:2011wm,Chen:2010bh,Li:2010ch,Castro:2010fd,Krishnan:2010pv,kerrcftstring,kerrcftsugra,kerrcftind,daCunha:2010jj,Wu:2009di,Huang:2010yg,Castro:2009jf,Guneratne:2012qp,ChangYoung:2012kd}. This correspondence and its extensions exploit the fact that extremal black holes exhibit a decoupled $AdS_2\times S^2$ near horizon topology \cite{Kunduri:2007vf}. This fact extends to the rotating charged \ks black hole of four dimensional low energy heterotic string theory \cite{Sen:1992ua} whose extremal near horizon geometry in conjunction with diffeomorphism invariance where employed in \cite{Ghezelbash:2009gf} to provide a microscopic derivation of its associated extremal black hole entropy. The non-geomertic hidden conformal symmetries of the \ks spacetime where explored in \cite{Ghezelbash:2012qn} and in conjunction with the extreme central charge, recycled to compute the non-extremal entropy. In this paper we add to the generality of these studies by computing the near-extremal \ks metric and by applying a combination of effective action approach and holography, account for the near-extremal \ks black hole entropy and temperature within one unified $AdS_2/CFT_1$ framework. 

As we will discover soon, the near-extremal \ks black hole also exhibits a decoupled near horizon geometry and is diffeomorphic to the extremal near horizon geometry of \cite{Ghezelbash:2009gf} via Poincar\'e coordinates. However, this diffeomorphism is singular on the asymptotic boundary of interest and implies an equivalence at the classical level, but differing quantum theories \cite{Compere:2012jk}. The near-extremal \ks geometry is characterized by a small but finite near-extremality parameter $\epsilon$, which is proportional to the Hawking temperature $\sim T_H$. This finite excitation, above extremality, allows for the computation of a properly normalized lowest Virasoro eigen-mode. The ability to compute a lowest Virasoro eigen-mode is a unique feature to the near-extremal case and is synonymous for the asymptotic symmetry group to include a proper $SL(2,\mathbb{R})$ subgroup allowing for the implementation of a traditional statistical (vs. thermal) Cardy formula. As a result we are able to determine the near-extremal \ks entropy and temperature without mixing quantitates derived separately at extremality or non-extremality. Finally, we determine the near-extremal entropy to be comprised of the ground state (extremal entropy) plus a non-degenerate excitation in terms of $\epsilon$\footnote{A feature, so far and to the best of our knowledge, only observed for the near-extremal Kerr spacetime \cite{Castro:2009jf,Compere:2012jk}.}.

Our computational construction mirrors that of \cite{Button:2010kg,ry}, where the near horizon isotropy is exploited to dimensionally reduce physics to two dimensions \emph{a la} Robinson and Wilczek \cite{robwill}. The idea is that most four dimensional near horizon geometries exhibit a sort of splitting:
\begin{align}
\label{eq:pbh}
ds^2=K_1(\theta)\left[ds_{(2)}^2+e^{-2\psi}\ell^2d\theta^2\right]+e^{-2\psi}\ell^2K_2(\theta)\left(d\phi+\mathcal{A}\right)^2
\end{align}
into two dimensional black hole $g^{(2)}_{\mu\nu}$, scalar/dilaton $\psi$ and $U(1)$ gauge field $\mathcal{A}=\A_\mu dx^\mu$. Mathematically $\psi$ may be treated as a matter field in the background of $g^{(2)}_{\mu\nu}$ and $\mathcal{A}$, however a semi-classical analysis of the scalar shows quantum gravitational effects due to its origin in the four dimensional parent black hole \eqref{eq:pbh}. We are then left with choosing appropriate boundary fall off conditions such that the two dimensional black hole and gauge field form an asymptotic $AdS_2$ configuration and then implementing an $AdS_2/CFT_1$ correspondence to study the thermodynamics of the near-extremal \ks black hole. 

The rest of the article is organized as follows: we discuss the relevant general \ks geometry and derive its near horizon near-extremal extension while defining all the relevant parameters and conventions in Section~\ref{sec:geo}. In Section~\ref{sec:qft} we perform a Robinson-Wilczek two dimensional reduction to the near horizon near-extremal extension and construct the effective action of the relevant two dimensional field content. Section~\ref{sec:asg} is devoted to the computation of the full asymptotic symmetry group using Lagrangian methods, the results of which are employed in Section~\ref{sec:adscft} to compute the near-extremal \ks thermodynamics. We conclude in Section~\ref{sec:con} with comments and discussion of our results.
\section{Near Horizon Near-Extremal \texorpdfstring{$Kerr$-$Sen$}{Kerr-Sen} Geometry}
\label{sec:geo}
The \ks black hole is a solution to four dimensional low energy heterotic string theory given by the line element:
\begin{align}
\label{eq:ksds}
\bs
ds^2=-\left(\frac{\Sigma\Delta}{\Pi}\right)d\tilde t^2+\Sigma\left(\frac{dr^2}{\Delta}+d\theta^2\right)
+\frac{\sin^2\theta}{\Sigma}\Pi\left(d\tilde \phi-\frac{2Mard\tilde t}{\Pi}\right)^2
\es
\end{align}
with
\begin{align}
\label{}
\notag\Sigma=&r(r+\varrho)+a^2\cos^2\theta &\Delta=&r(r+\varrho)-2Mr+a^2\\
\notag\Pi=&\left(r(r+\varrho)+a^2\right)^2-\Delta a^2\sin^2\theta&&\\
&\notag\text{mass} &M=&m\cosh^2\alpha\\
&\notag\text{angular-momentum} &J=&ma\cosh^2\alpha\\
&\notag\text{charge} &Q^2=&\varrho M=2m\sinh^2\alpha,
\end{align}
where $\alpha$ is an arbitrary parameter (of the solution generating method) and $m$ and $a$ are the mass and angular momentum per unit mass of the resulting $Kerr$ black hole in the limit when $\alpha=0$ \cite{Sen:1992ua}. We will be working in units where Newton's constant and the speed of light obey $G=c=1$, except when it may be useful to reintroduce them for clarification or to reinforce a specific concept. The horizons of \eqr{eq:ksds} are located at:
\begin{align}
\label{eq:rp}
r_\pm=M-\frac{\varrho}{2}\pm\sqrt{\left(M-\frac{\varrho}{2}\right)^2-a^2}
\end{align}
where the extremal limit is defined by
\begin{align}
\label{eq:exlim}
J=M\left(M-\frac{\varrho}{2}\right)=\tilde M^2
\end{align}
from which we obtain the non-extremal and extremal Bekenstein-Hawking entropy and Hawking temperature \cite{Sen:1992ua}:
\begin{align}
\label{eq:ent}
\bs
S(r_+)=&2\pi M\left[\left(M-\frac{\varrho}{2}\right)+\sqrt{\left(M-\frac{\varrho}{2}\right)^2-\frac{J^2}{M^2}}\right],\\
S_{ext}=&2\pi J~\text{and}\\
T_H=&\frac{r_+-\left(M-\frac{\varrho}{2}\right)}{4\pi M r_+}.
\es
\end{align}
We will not be concerned with the other four dimensional field content of this theory since we are only interested in gravitational thermodynamics and the other four dimensional fields do not in general contribute to the center of the asymptotic symmetry group \cite{Ghezelbash:2009gf,Hartman:2008pb,Compere:2009dp}.

The near horizon extremal \ks geometry ($NHEKS$) is singled out via the coordinate transformations:
\begin{align}
\label{eq:exct}
r=\left(M-\frac{\varrho}{2}\right)+\lambda U,~\tilde\phi=\phi+\frac{t}{\lambda2M}~\text{and}~\tilde t=\frac{t}{\lambda}
\end{align}
in the limit as $\lambda\to0$. Substituting the above into \eqr{eq:ksds} and taking the limit $\lambda\to0$ yields:
\begin{align}
\label{eq:nheks}
\bs
ds_{NHEKS}^2=\frac{2 M+\varrho+(2M-\varrho)\cos^2\theta}{4M}\left(-\frac{U^2}{2M\left(M-\frac{\varrho}{2}\right)}dt^2+\frac{2M\left(M-\frac{\varrho}{2}\right)}{U^2}dU^2\right.\\
\left.+2M\left(M-\frac{\varrho}{2}\right)d\theta^2\right)+\frac{4M\left(2M\left(M-\frac{\varrho}{2}\right)\right)\sin^2\theta}{2 M+\varrho+(2M-\varrho)\cos^2\theta}\left(d\phi+\frac{U}{2M\left(M-\frac{\varrho}{2}\right)}dt\right)^2.
\es
\end{align}
The above line element has an obvious $AdS_2\times S^2$ topology with $AdS_2$ scale 
\begin{align}
\label{eq:ads2}
   \ell^2=2Mr_+=2M\left(M-\frac{\varrho}{2}\right)=2J.
\end{align}
We seek a modification of the above line element that results in a finite excitation above extremality, i.e.
\begin{align}
\label{eq:eaex}
\tilde M=\sqrt{J}+\mathcal{E},~\text{where}~\mathcal{E}\ll\sqrt{J}.
\end{align}
To achieve this, we define the near-extremality parameter $\epsilon$ such that\footnote{The relation between the excitation energy and near-extremal parameter is given by $\mc{E}=\frac{M^2\epsilon^2\lambda^2}{4J^{3/2}}$.}:
\begin{align}
\label{}
\bs
\epsilon\lambda=&r_+-\left(M-\frac{\varrho}{2}\right)\\
&\text{and}\\
\ell^2=&2\left(J+\sqrt{4J^{3/2}\mc{E}}\right)+\mc{O}(\epsilon\lambda)^2.
\es
\end{align}
Next, rewriting the radial part of \eqr{eq:exct}, the horizons and angular momentum per mass as:
\begin{align}
\label{eq:nexct}
r=\left(\frac{\ell^2}{2M}\right)+\lambda (U-\epsilon),~r_+=\frac{\ell^2}{2M},~r_-=\frac{\ell^2}{2M}-2\lambda\epsilon,~a=\sqrt{\frac{\ell^2}{2M}\left(\left(M-\frac{\varrho}{2}\right)-\lambda\epsilon\right)}
\end{align}
and using
\begin{align}
\label{eq:nextdsr}
\bs
\Delta=&\left(U^2-\epsilon^2\right)\lambda^2\\
\Sigma=&\frac{\ell^2\left(\ell^2+2M\varrho+M(2M-\varrho)\cos^2\theta\right)}{4M^2}+\mc{O}(\lambda)\\
\Pi=&\ell^4\left(1+\frac{4MU\lambda}{\ell^2}\right)+\mc{O}(\lambda)^2
\es
\end{align}
in \eqr{eq:ksds}, while holding both $\epsilon$ and $\ell^2$ fixed in the limit as $\lambda\to0$ yields the near horizon near-extremal \ks ($NHNEKS$) geometry:
\begin{align}
\label{eq:nhneks}
\bs
ds_{NHEKS}^2=&\frac{\ell^2+2M\varrho+M(2M-\varrho)\cos^2\theta}{4M^2}\left(-\frac{U^2-\epsilon^2}{\ell^2}dt^2+\frac{\ell^2}{U^2-\epsilon^2}dU^2+\ell^2d\theta^2\right)\\
&+\frac{4M^2\ell^2\sin^2\theta}{\ell^2+2M\varrho+M(2M-\varrho)\cos^2\theta}\left(d\phi+\frac{U}{\ell^2}dt\right)^2.
\es
\end{align}
The above line element still exhibits a global $AdS_2\times S^2$ topology. However, it now has a horizon located at $U=\epsilon$ and a finite temperature $T_{H}=\epsilon/(2\pi\ell^2)$. 

As mentioned in the Introduction, both near horizon geometries, \eqr{eq:nheks} and \eqr{eq:nhneks}, are classically diffeomorphic as seen by the form of their Ricci scalars:
\begin{align}
\label{eq:rscc}
R=
\begin{cases}
\frac{64M\varrho^2 \sin^2\theta}{\ell^2(\cos(2 \theta) (2M-\varrho )+6 M+\varrho )^3}&NHEKS\\
~\\
\frac{64M^3 \sin^2\theta\left(\ell^2(\varrho -2 M)+2 M \left(2 M^2-2 M \varrho +\varrho^2\right)\right)}{\ell^2 (2\ell^2+M \cos(2 \theta ) (2 M-\varrho )+M (2 M+3 \varrho ))^3}&NHNEKS
\end{cases},
\end{align}
which are identical in the extreme limit when $\ell^2=2M(M-\varrho/2)$\footnote{This statement extends to all the Gauss-Bonnet invariants.}. However, the finite horizon and temperature in terms of $\epsilon$, for the near-extremal case, allow for interesting and non trivial additions to the quantum geometric analysis of \eqr{eq:nhneks} and its thermodynamics.
\section{Quantum Fields in \texorpdfstring{$NHNEKS$}{NHNEKS}}
\label{sec:qft}
As discussed in the introduction, our goal is to perform a semi-classical analysis of the two dimensional scalar/dilaton field with respect to the Kaluza-Klein field content of \eqr{eq:nhneks}. We know the quantum effective action of a minimally coupled scalar has the form:
\begin{align}
\label{eq:seffa}
S_{eff}\sim\frac{\bar\beta^\psi}{16\pi}\int d^2x\sqrt{-g^{(2)}}R^{(2)}\frac{1}{\square_{g^{(2)}}}R^{(2)}+\cdots,
\end{align}
where  $\bar\beta^\psi=\frac{const}{G}$ is the Weyl anomaly coefficient \cite{tseytlin89,polyak}. We will determine the full effective theory and value of $\bar\beta^\psi$ to $s$-wave. This is a sensible approximation since $\psi$ is gravitational in origin and as such, should be real and unit-less. Furthermore, most of the gravitational dynamics seem to be contained in this region of approximation \cite{strom1}\footnote{In \cite{Button:2010kg} it was shown that $\varphi_{lm}$ decays exponentially fast in time by analyzing the asymptotic behavior of its field equation for higher orders in $l$ and $m$.}.

To single out the relevant two dimensional near horizon theory, we begin by considering a four dimensional massless free scalar field in the background of \eqr{eq:nhneks}:
\begin{align}
\label{eq:freescalar4}
\bs
S_{free}=&\frac12\int d^4x\sqrt{-g}g^{\mu\nu}\partial_{\mu}\varphi\partial_\nu\varphi\\
=&-\frac12\int d^4x\,\varphi\left[\partial_\mu\left(\sqrt{-g}g^{\mu\nu}\partial_\nu\right)\right]\varphi\\
=&-\frac12\int d^4x\,\varphi\left[\left( -\ell^2\sin{\theta}\frac{\ell^2}{U^2-\epsilon^2}\partial_t^2\right)+\right.\\
&\ell^2\sin{\theta}\partial_U\left(\frac{U^2-\epsilon^2}{\ell^2}\partial_U\right)+\partial_\theta\left( \sin{\theta}\partial_\theta\right)+\\
&\left( \left\{-\ell^2\sin{\theta}\left(\frac{U}{\ell^2}\right)^2\frac{\ell^2}{U^2-\epsilon^2}+\frac{\left(2 \ell^2+M (2 M+3 \varrho )+M  (2 M-\varrho )\cos (2 \theta )\right)^2}{64M^4\sin\theta}\right\}\partial_\phi^2\right)+\\
&\left.2\left( \ell^2\sin{\theta}\frac{U}{\ell^2}\frac{\ell^2}{U^2-\epsilon^2}\right)\partial_t\partial_\phi\right]\varphi.
\es
\end{align}
In order to reduce the above to a two dimensional theory, we must integrate away angular degrees of freedom. To this end, we expand $\varphi$ in terms of spherical harmonics
\begin{align}
\label{eq:sphdecom}
\varphi(t,U,\theta,\phi)=\sum_{lm}\varphi_{lm}(t,U)Y\indices{_l^m}(\theta,\phi),
\end{align}
and transform to tortoise coordinates defined by $\frac{dU^{*}}{dU}=\frac{\ell^2}{U^2-\epsilon^2}$. Furthermore, the resulting two dimensional theory is much simplified by considering the region $U\sim\epsilon$, since mixing and potential terms ($\sim l(l+1)\ldots$) are weighted by a factor of $f(U(U^*))\sim e^{2\kappa U^*}$, which vanish exponentially fast as $U\to \epsilon$. This leaves us with the remnant functional:
\begin{align}
\label{eq:nhapw}
\bs
S_{free}=&-\frac{\ell^2}{2}\int d^2x\;\varphi^{*}_{lm}\left[-\frac{\ell^2}{U^2-\epsilon^2}\left(\partial_t-im\frac{U}{\ell^2}\right)^2+\partial_U\frac{U^2-\epsilon^2}{\ell^2}\partial_U\right]\varphi_{lm}\\
=&-\frac{\ell^2}{2}\int d^2x\;\varphi^{*}_{lm}\left[-\frac{1}{f(U)}\left(\partial_t-im\A_t\right)^2+\partial_Uf(U)\partial_U\right]\varphi_{lm}\\
=&-\frac{\ell^2}{2}\int d^2x\;\varphi^{*}_{lm}D_{\mu}\left[\sqrt{-g^{(2)}}g^{\mu\nu}_{(2)}D_{\nu}\right]\varphi_{lm},
\es
\end{align}
where $D_\mu=\partial_\mu-im\A_\mu$ is the gauge covariant derivative with $U(1)$ two dimensional complex scalar charge $e=m$ and we have introduced the Robinson-Wilczek two dimensional analogue (RW2DA) fields:
\begin{align}
\label{eq:rwdaf}
g^{(2)}_{\mu\nu}=&\left(\begin{array}{cc}-f(U) & 0 \\0 & \frac{1}{f(U)}\end{array}\right)&\A=&\A_tdt\\
f(U)=&\frac{U^2-\epsilon^2}{\ell^2}&\A_t=&\frac{U}{\ell^2}.
\end{align}
The quantum effective functional of \eqr{eq:nhapw}, to $s$-wave $\varphi_{00}=\sqrt{\frac{6}{G}}\psi$ for unitless\footnote{The factor of $\sqrt{6}$ is chosen to coincide with the normalization of \eqr{eq:seffa} for the gravitational sector of the effective action.} $\psi$, is obtained via path integrating over $\psi$, which amounts to a zeta-function regularization of the functional determinant of $D_{\mu}\left[\sqrt{-g^{(2)}}g^{\mu\nu}_{(2)}D_{\nu}\right]$. It is comprised of the two parts \cite{Leutwyler:1984nd,isowill}:
\begin{align}\label{eq:nh2pcft}
\Gamma=&\Gamma_{grav}+\Gamma_{U(1)},
\end{align}
where
\begin{align}
\label{eq:twoeffac}
\bs
\Gamma_{grav}=&\frac{\ell^2}{16\pi}\int d^2x\sqrt{-g^{(2)}}R^{(2)}\frac{1}{\square_{g^{(2)}}}R^{(2)}~\mbox{and}\\
\Gamma_{U(1)}=&\frac{3 e^2 \ell^2}{\pi }\int \mc{F}\frac{1}{\square_{g^{(2)}}}\mc{F}
\es
\end{align}
in  concurrence with \eqr{eq:seffa} for $\bar\beta^\psi=\frac{\ell^2}{G}$. To restore locality in the above functionals, we introduce auxiliary (dilation and axion) scalars $\Phi$ and $B$ such that:
\begin{align}\label{eq:afeqm}
\square_{g^{(2)}} \Phi=R^{(2)}~\mbox{and}~\square_{g^{(2)}} B=\epsilon^{\mu\nu}\partial_\mu \A_\nu,
\end{align}
which in terms of general $f(U)$ and $\A_t(U)$ read:
\begin{align}
\label{eq:afsol}
\bs
-\frac{1}{f(U)}\partial_t^2\Phi+\partial_Uf(U)\partial_U\Phi=&R^{(2)}\\
-\frac{1}{f(U)}\partial_t^2B+\partial_Uf(U)\partial_UB=&\mc{F}
\es
\end{align}
and exhibit the general solutions:
\begin{align}
\label{eq:afsol}
\bs
\Phi(t,U)=&\alpha_1 t+\int dU\frac{\alpha_2-f'(U)}{f(U)}\\
B(t,U)=&\beta_1 t+\int dU\frac{\beta_2-\A_t(U)}{f(U)},
\es
\end{align}
where $\alpha_i$ and $\beta_i$ are integration constants. Next, to obtain a local action we make use of \eqr{eq:afeqm} to transform \eqr{eq:nh2pcft} into our final Liouville type near horizon $CFT$, yielding:
\begin{align}\label{eq:nhfinalcft}
\bs
S_{NHCFT}=&\frac{\ell^2}{16\pi}\int d^2x\sqrt{-g^{(2)}}\left\{-\Phi\square_{g^{(2)}}\Phi+2\Phi R^{(2)}\right\}\\
&+\frac{3 e^2 \ell^2}{\pi}\int d^2x\sqrt{-g^{(2)}}\left\{-B\square_{g^{(2)}}B+2B \left(\frac{\epsilon^{\mu\nu}}{\sqrt{-g^{(2)}}}\right)\partial_\mu A_\nu\right\}.
\es
\end{align}
Further, for use in the section below, we define the asymptotic (large $U$) boundary fields:
\begin{align}\label{eq:as2dwa}
g^{(0)}_{\mu\nu}=&
\left(
\begin{array}{cc}
-\frac{U^2}{\ell^2}+\frac{\epsilon^2}{\ell^2}+\mc{O}\left(\frac{1}{U}\right)^3& 0 \\
 0 & \frac{\ell^2}{U^2}+\mathcal{O}\left(\frac{1}{U}\right)^3
\end{array}
\right)\\
\label{eq:asgf}
\mathcal{A}\indices{^{(0)}_t}=&\frac{U}{\ell^2}+\mathcal{O}\left(\frac{1}{U}\right)^3,
\end{align}
which form an $AdS_2$ configuration with scalar curvature $R^{(2)}=-\frac{2}{\ell^2}+O\left(\frac{1}{U}\right)$. A substitution of these fields into \eqr{eq:afsol} allows us to define the boundary auxiliary scalars $\Phi^{(0)}=\Phi(g^{(0)},\mathcal{A}^{(0)})$ and $B^{(0)}=B(g^{(0)},\mathcal{A}^{(0)})$:
\begin{align}
\label{eq:bfasc}
\bs
\Phi^{(0)}(t,U)=&\alpha_1t+2\ln\left(\frac{1}{U}\right)-\frac{\alpha_2\ell ^2}{U}+\frac{\epsilon
   ^2}{U^2}-\frac{\alpha_2 \ell ^2 \epsilon ^2}{3U^3}+\mc{O}\left(\frac{1}{U}\right)^4\\
B^{(0)}(t,U)=&\beta_1t+\frac{1-\beta_2\ell ^2}{U}-\frac{\epsilon^2\left(\beta_2\ell ^2-1\right)}{3U^3}+\mc{O}\left(\frac{1}{U}\right)^4,
\es
\end{align}
for still to be determined constants $\alpha_i$ and $\beta_i$.
\section{Asymptotic Symmetries}
\label{sec:asg}
In this section we will turn our attention to the quantum asymptotic symmetry group of the RW2DA fields. Given the  boundary fields \eqr{eq:as2dwa} and \eqr{eq:asgf} we begin by imposing the following metric, gauge field and scalar field fall of conditions:
\begin{align}\label{eq:mbc}
\delta g_{\mu\nu}=
\left(
\begin{array}{cc}
    \mathcal{O}\left(\frac{1}{U}\right)^3&
   \mathcal{O}\left(\frac{1}{U}\right)^0 \\
 \mathcal{O}\left(\frac{1}{U}\right)^0 &
\mathcal{O}\left(U\right) 
\end{array}
\right),~\delta \mathcal{A}=\mathcal{O}\left(\frac{1}{U}\right)^0,~\delta\Phi=\mathcal{O}\left(\frac{1}{U}\right)^0~\mbox{and}~\delta B=\mathcal{O}\left(\frac{1}{U}\right)^0.
\end{align}
A set of diffeomorphisms preserving the above configuration is given by:
\begin{align}\label{eq:dpr}
\chi=-C_1\frac{U^2 \xi(t)}{U^2-\epsilon^2}\partial_t+C_2U\xi'(t)\partial_U,
\end{align}
where $\xi(t)$ is an arbitrary function and $C_i$ are arbitrary normalization constants. The variation of the gauge field under the above diffeomorphism is given by $\delta_\chi \mathcal{A}=\mathcal{O}\left(\frac{1}{U}\right)^0$ and thus, $\delta_\chi$ is trivially elevated to a total symmetry 
\begin{align}
\label{eq:totsym}
\delta_{\chi}\to\delta_{\chi+\Lambda}
\end{align}
of the action. Switching to conformal light cone coordinates
\begin{align}
x^\pm=t\pm U^*
\end{align}
and transforming the above diffeomorphism, we obtain the components:
\begin{align}
\label{eq:lcdiff}
\chi^\pm=\frac{U(U^*)\left(-C_1U(U^*) \xi(x^+,x^-)\pm C_2\ell^2 \xi'(x^+,x^-)\right)}{U(U^*)^2-\epsilon ^2},
\end{align}
which are well behaved on the asymptotic boundary. 

The corresponding quantum generator is defined by the conserved charge:
\begin{align}
\label{eq:ccharge}
\mathcal{Q}(\chi)=\int dx^\mu\left\langle T_{\mu\nu}\right\rangle\chi^\nu,
\end{align}
where $\left\langle T_{\mu\nu}\right\rangle$ is the energy momentum tensor (EMT) of \eqr{eq:nhfinalcft} defined by:
\begin{align}
\label{eq:emt}
\bs
\left\langle T_{\mu\nu}\right\rangle=&\frac{2}{\sqrt{-g^{(2)}}}\frac{\delta S_{NHCFT}}{\delta g_{(2)}^{\mu\nu}}\\
=&\frac{\ell^2}{8\pi}\left\{\partial_\mu\Phi\partial_\nu\Phi-2\nabla_\mu\partial_\nu\Phi+g^{(2)}_{\mu\nu}\left[2R^{(2)}-\frac12\nabla_\alpha\Phi\nabla^\alpha\Phi\right]\right\}\\
&+\frac{6 e^2 \ell^2}{\pi}\left\{\partial_\mu B\partial_\nu B-\frac12g_{\mu\nu}\partial_\alpha B\partial^\alpha B\right\}\\
&\text{and}\\
\left\langle J^{\mu}\right\rangle=&\frac{1}{\sqrt{-g^{(2)}}}\frac{\delta S_{NHCFT}}{\delta \mathcal{A}_\mu}=\frac{6 e^2 \ell^2}{\pi}\frac{1}{\sqrt{-g^{(2)}}}\epsilon^{\mu\nu}\partial_\nu B
\es
\end{align}
is the $U(1)$ current, listed for completeness. Substituting the general solutions \eqr{eq:afsol} into \eqr{eq:emt} and adopting modified Unruh Vacuum boundary conditions (MUBC) \cite{unruh}
\begin{align}
\label{eq:ubc}
\begin{cases}
\left\langle T_{++}\right\rangle=\left\langle J_{+}\right\rangle=0&U\rightarrow\infty,~\ell\rightarrow\infty\\
\left\langle T_{--}\right\rangle=\left\langle J_{-}\right\rangle=0&U\rightarrow \epsilon
\end{cases},
\end{align}
we obtain the general behavior:
\begin{align}
\label{}
\begin{cases}
f(U)=0&U\to \epsilon\\
f(U)=\A_t(U)=0&\ell\to \infty\\
\end{cases}.
\end{align}
These conditions allow us to determine the integration constants $\alpha_i$ and $\beta_i$:
\begin{align}
\label{}
\bs
\alpha_1=&-\alpha_2=\frac12f'(\epsilon) \\
\beta_1=&-\beta_2=\frac12\A_t(\epsilon) 
\es
\end{align}
thus, specifying both the EMT and $U(1)$ current. A straightforward calculation reveals that the EMT exhibits a Weyl (trace) anomaly given by:
\begin{align}
\label{eq:tra}
\left\langle T\indices{_\mu^\mu}\right\rangle=-\frac{\bar\beta^\psi}{4\pi}R^{(2)},
\end{align}
which determines the central charge via \cite{cft}:
\begin{align}
\label{eq:center1}
\frac{c}{24\pi}=\frac{\bar\beta^\psi}{4\pi}\Rightarrow c=6\ell^2=12\left(J+\sqrt{4J^{3/2}\mc{E}}\right)+\mc{O}(\epsilon^2).
\end{align}
A unique advantage of the MUBC is that at the asymptotic boundary of interest (and to $\mathcal{O}(\frac{1}{\ell})^2$, which will be denoted by the single limit $x^+\to\infty$) the EMT is dominated by one holomorphic component $\left\langle T_{--}\right\rangle$. Expanding this component in terms of the boundary fields \eqr{eq:as2dwa} and \eqr{eq:asgf} (or \eqr{eq:bfasc}) and computing its response to a total symmetry we get:
\begin{align}
\begin{cases}
\delta_{\chi^-+\Lambda}\left\langle T_{--}\right\rangle=\chi^-\left\langle T_{--}\right\rangle'+2\left\langle T_{--}\right\rangle\left(\chi^-\right)'+\frac{c}{24\pi}\left(\chi^-\right)'''+\mathcal{O}\left(\left(\frac{1}{U}\right)^3\right)\\
\delta_{\chi^-+\Lambda}\left\langle J_{-}\right\rangle=\mathcal{O}\left(\left(\frac{1}{U}\right)^3\right)
\end{cases},
\end{align}
while keeping in mind that $\delta_{\xi^{\pm}+\Lambda}g_{\pm\mp}=\xi^{\pm}\partial_{\pm}g_{\pm\mp}+g_{\pm\mp}\partial_{\pm}\xi^{\pm}$ $\Rightarrow$ $\delta_{\xi^{\pm}+\Lambda}f=\xi^{\pm}\partial_{\pm}f+f\partial_{\pm}\xi^{\pm}$ and $\delta_{\xi^{\pm}+\Lambda}\mc{A}_{\pm}=\xi^{\pm}\partial_{\pm}\mc{A}_{\pm}+\mc{A}_{\pm}\partial_{\pm}\xi^{\pm}+\partial_{\pm}\Lambda$\footnote{We need only consider trivial gauge transformations as explained in the paragraph above \eqr{eq:totsym}.}. This shows that $\left\langle T_{--}\right\rangle$ transforms asymptotically as the EMT of a one dimensional $CFT$. Finally we will compute the asymptotic charge algebra by compactifying the $x^-$ coordinate to a circle from $0\to2\pi\ell^2/\epsilon$ and introducing the asymptotic conserved charge:
\begin{align}
\label{eq:asycharge}
\mathcal{Q}_n=\lim_{x^+\to\infty}\int dx^\mu\left\langle T_{\mu\nu}\right\rangle\chi^\nu_n,
\end{align}
where $\xi(x^+,x^-)$ has been replaced by circle diffieomorphisms $\frac{e^{-in(\ell^2/\epsilon)x^\pm}}{\ell^2/\epsilon}$ in \eqr{eq:lcdiff} and the $C_i$ are fixed by requiring the $\chi^-_n$ to form an asymptotic centerless Witt or $Diff(S^1)$ subalgebra:
\begin{align}
i\left\{\chi^-_m,\chi^-_n\right\}=(m-n)\chi^-_{m+n}.
\end{align}
Next, calculating the response of $\mathcal{Q}_n$ to a total symmetry we have:
\begin{align}
\label{eq:ca}
\delta_{\chi^-_m+\Lambda}\Q_n=\left[\Q_m,\Q_n\right]=(m-n)\Q_n+\frac{c}{12}m\left(m^2-1\right)\delta_{m+n,0},
\end{align}
which shows that the asymptotic quantum generators form a centrally extended Virasoro algebra with center \eqr{eq:center1} and computable non-zero lowest eigen-mode, $\Q_0=\frac{\left(J+\sqrt{4J^{3/2}\mc{E}}\right)}{2}+\mc{O}(\epsilon^2)$.
\section{\texorpdfstring{$AdS_2/CFT_1$}{ADS} and Near-Exremal \texorpdfstring{$Kerr$-$Sen$}{KS} Thermodynamics}
\label{sec:adscft}
By computing the asymptotic symmetry group of the RW2DA fields, we have shown that the \neks spacetime is holographically dual to a $CFT$ with center
\begin{align}\label{eq:cre}
c&=12\left(J+\sqrt{4J^{3/2}\mc{E}}\right)+\mc{O}(\epsilon^2)
\end{align}
and lowest Virasoro eigen-mode 
\begin{align}
\Q_0&=\frac{\left(J+\sqrt{4J^{3/2}\mc{E}}\right)}{2}+\mc{O}(\epsilon^2).
\end{align}
We are now in a position to employ the above results within a traditional statistical Cardy formula, yielding the near-extremal \ks entropy:
\begin{align}\label{eq:nexent}
S_{near-ext}=2\pi\sqrt{\frac{c\Q_0}{6}}=2\pi\left(J+\sqrt{4J^{3/2}\mc{E}}\right)+\mc{O}(\epsilon^2).
\end{align}
The near-etremal entropy has a peculiar form:
\begin{align}\label{eq:nexent}
S_{near-ext}=S_{ground}+S_{1^{st}~excited-state},
\end{align}
where $S_{ground}=2\pi J$ is the extremal \ks entropy and $S_{1^{st}~excited-state}=2\pi\sqrt{\frac{12J h}{6}}$, where $h=2\sqrt{J}\mc{E}$, is the excitation entropy just above extremality (the above result and $h$ will be discussed more in the conclusion).

An additional interesting question concerns the \neks horizon temperature and if it may be derived from our constructed correspondence. To answer this question, we will focus on the gravitational part of \eqr{eq:nhfinalcft}, given by:
\begin{align}\label{eq:nhlcftgrav}
S_{grav}=&\frac{\ell^2}{16\pi}\int d^2x\sqrt{-g^{(2)}}\left\{-\Phi\square_{g^{(2)}}\Phi+2\Phi R^{(2)}\right\}
\end{align}
and with energy momentum:
\begin{align}
\label{eq:emtgra}
\bs
\left\langle T_{\mu\nu}\right\rangle=&\frac{2}{\sqrt{-g^{(2)}}}\frac{\delta S_{NHCFT}}{\delta g\indices{^{(2)}^\mu^\nu}}\\
=&\frac{\ell^2}{8\pi}\left\{\partial_\mu\Phi\partial_\nu\Phi-2\nabla_\mu\partial_\nu\Phi+g\indices{^{(2)}_\mu_\nu}\left[2R^{(2)}-\frac12\nabla_\alpha\Phi\nabla^\alpha\Phi\right]\right\}.
\es
\end{align}
Next, repeating steps \eqr{eq:emt} and \eqr{eq:ubc}, but focusing on the horizon limit $U\to \epsilon$ rather than the $x^+\to\infty$ boundary, we obtain the single holomorphic component,
\begin{align}
\label{eq:hhf}
\left\langle T_{++}\right\rangle=-\frac{\ell^2}{32 \pi }f'\left(\epsilon\right)^2,
\end{align}
which is precisely the value of the Hawking Flux of the \neks black hole weighted by the central charge \eqr{eq:center1}. This allows us to extract the Hawking temperature \cite{Xu:2006tq,Jinwu,caldarelli:1999xj} via the standard identifications:
\begin{align}
\left\langle T_{++}\right\rangle=cHF=-c\frac{\pi}{12}\left(T_H\right)^2\Rightarrow T_H=\frac{f'\left(\epsilon\right)}{4 \pi }.
\end{align}
This result tells us that the $AdS_2/CFT_1$ correspondence constructed here intrinsically contains information about both black hole entropy and temperature, though the temperature is only extractible from prior knowledge of the central extension of the asymptotic symmetry group.
\section{Conclusion}
\label{sec:con}
We have analyzed black hole thermodynamics of the \neks spacetime by constructing a near horizon effective action of its RW2DA fields and computing the resulting quantum asymptotic symmetry group. This implies a $AdS_2/CFT_1$ correspondence in the near horizon of near-extremal \ks metric with:
\begin{align}\label{eq:concft}
\bs
c=&12\left(J+\sqrt{4J^{3/2}\mc{E}}\right)+\mc{O}(\epsilon^2),\\
\Q_0=&\frac{\left(J+\sqrt{4J^{3/2}\mc{E}}\right)}{2}+\mc{O}(\epsilon^2)~\text{and}\\
S_{near-ext}=&2\pi\left(J+\sqrt{4J^{3/2}\mc{E}}\right)+\mc{O}(\epsilon^2).
\es
\end{align}
\eqr{eq:ubc}
An obvious question that arises from the above, is how do these results compare or relate to the traditional $Kerr/CFT$ correspondence of the \ks black hole. In the $Kerr/CFT$ picture, the extremal \ks black hole is shown to be dual to a chiral $CFT$ with centers  and temperatures \cite{Ghezelbash:2009gf}. 
\begin{align}
\label{}
c_L=&12J,&c_R=&12J\\
T_L=&\frac{1}{2\pi}~\text{and}&T_R=&0
\end{align}
thus, yielding the extremal entropy within a thermal Cardy formula:
\begin{align}
\label{}
S_{ext}=\frac{\pi^2}{3}(c_LT_L+c_RT_R)=2\pi J.
\end{align}
The departure from extremality should be accounted for in the entropy by contributions of the right sector of the chiral $CFT$. This is evident in the non-extremal case \cite{Ghezelbash:2012qn}, but requires a non geometric analysis and incorporating results derived separately at extremality. The chiral structure seems to be missing in our results of \eqr{eq:concft}, which is rooted in our implementation of the MUBC \eqr{eq:ubc}, which singles out a specific homlomorphic component of the quantum EMT \eqr{eq:emt} at the asymptotic boundary. One possible interpretation is that this singling out forces both chiral contributions into one $c$ and $\Q_0$. Indeed, as mentioned in Section~\ref{sec:adscft}, looking at the near-extremal entropy we see it has the form:
\begin{align}
\label{}
S_{near-ext}=&2\pi J+2\pi\sqrt{\frac{c_R h_R}{6}}+\mc{O}(\epsilon^2)\\
=&S_{ground}+S_{1^{st}~excited-state}+\cdots,
\end{align}
 where $h_R=2\sqrt{J}\mc{E}$. In other words, the ground state entropy is completely specified by the left sector and the first excited state is specified by the right central charge and weight $h_R$ within a more traditional Cardy formula. This scenario, though plausible and interesting, requires more follow up calculations before a definitive statement can be made.
 
Another interesting feature is the dependence of the near extremality parameter $\epsilon$ in our construction. Taking the limit back to extremality, $\epsilon\to0$, would result in several divergences during intermediary computations. Though the final results of the asymptotic symmetry group are well defined, it is not clear how to extend the full calculation to this limit. However, there are still ample spacetime testing beds beyond $Kerr$ which lack a near-extremality analysis of their asymptotic symmetry groups. Perhaps application of the developed techniques in this article to a greater diversity of such spacetimes could help to furnish the correspondence and better answer some of the questions posed above. 
\acknowledgments
We thank Vincent Rodgers and Steven Carlip for enlightening discussions. L.R. is grateful to the University of Iowa and NASA Goddard Space Flight Center for their hospitality. This work is supported in part by the HHMI Undergraduate Science Education Award 52006298 and the Grinnell College Academic Affairs' CSFS and MAP programs.
\appendix


\bibliographystyle{JHEP}
\bibliography{cftgr}



\end{document}